# Frustration-induced diffusive scattering anomaly and dimension change in FeGe$_2$

Yaokun Su[1], Hillary L. Smith[2], Matthew B. Stone[3], Douglas L. Abernathy[3], Mark D. Lumsden[3], Carl P. Adams[4], and Chen Li[1,5*]

1. Materials Science and Engineering, University of California, Riverside, Riverside, CA 92521, USA
2. Physics and Astronomy, Swarthmore College, Swarthmore, PA 19081, USA
3. Neutron Scattering Division, Oak Ridge National Laboratory, Oak Ridge, TN 37831, USA
4. Department of Physics, St. Francis Xavier University, Antigonish, Nova Scotia B2G 2W5, Canada
5. Mechanical Engineering, University of California, Riverside, Riverside, CA 92521, USA

*chenli@ucr.edu

**Abstract**

Magnetic frustration, arising from the competition of exchange interactions, has received great attention because of its relevance to exotic quantum phenomena in materials. In the current work, we report an unusual checkerboard-shaped scattering anomaly in FeGe$_2$, far from the known incommensurate magnetic satellite peaks, for the first time by inelastic neutron scattering. More surprisingly, such phenomenon appears as spin dynamics at low temperature, but it becomes prominent above Néel transition as elastic scattering. A new model Hamiltonian that includes an intraplane next-nearest neighbor was proposed and attributes such anomaly to the near-perfect magnetic frustration and the emergence of unexpected two-dimensional magnetic order in the quasi-one-dimensional FeGe$_2$.

Magnetic frustration has attracted interest due to its relation to novel phases including quantum spin liquids, spin and electronic nematic phases and unconventional superconductivity [1–5]. Generally, the magnetic frustration arises with special geometry of lattice, but it can also be achieved as a consequence of the competition between different pair antiferromagnetic (AFM) interactions.

In recent years, AFM materials have been widely studied due to their robustness again disturbance and potential applications in high density data storage [6], resistive switching [7,8] and spintronics [9,10]. As a germanium-based AFM intermetallic, FeGe$_2$ has been explored by numerous experimental and theoretical studies for its complex magnetism [11–17]. Magnetic excitations have been measured in FeGe$_2$ [18,19], revealing a large anisotropy and an overdamped feature of the spin wave. A NN Heisenberg model was proposed with $SJ_c$ = 136 meV and $SJ_1$ = -8.8 meV, where S is the on-site spin magnitude and Js are the exchange constants. However, the measurements were highly restricted by the instruments used, and a detailed examination of its magnetic excitations throughout the full Brillouin zone is still missing.

FeGe$_2$ has the same body-centered tetragonal crystal structure [20] (Fig. S1) as $\theta$-phase Al$_2$Cu (space group *I4/mcm*). It exhibits two zero-field magnetic phase transitions on heating [17,21]: one first-order transition from a commensurate AFM state to an incommensurate spin-density-wave state at 263 K, and another second-order Néel transition from the incommensurate state to paramagnetic phase at 289 K. The ordering wavevector changes from $(2\pi/a)[1, 0, 0]$ for the commensurate state to $(2\pi/a)[1+\delta, 0, 0]$ for the incommensurate state, where $\delta$ varies from 0 to 0.05. Along the *c* axis, the nearest neighbor (NN) distance between Fe atoms is 2.478 Å, which is close to that of elemental Fe (2.482 Å), such that ferromagnetic (FM) exchange interaction $J_c$ is expected to be strong. In the *a-b* plane, only a weak NN AFM exchange interaction $J_1$ has been considered previously [18,19]. While the magnetic moments were known to lay in the basal plane with a value of 1.2 $\mu_B$ per Fe atom, their exact orientation is still unclear [21].

Here we provide a combined scattering and computational study of FeGe$_2$ in examination of a portion of the magnetic spectrum that was not previously identified. We use inelastic neutron scattering (INS) to examine the lattice and magnetic dynamics of FeGe$_2$, revealing an unusual checkerboard-shaped scattering anomaly. Such anomaly appears in dynamic part at low temperature and turns into diffusive intensity above Néel transition. An additional intraplane next-nearest neighbor (NNN) interaction J$_2$ is found to be necessary to fully describe the spin wave spectrum. Exchange parameters calculated from atomistic simulations and the new model including J$_2$ suggest the near-perfect magnetic frustration in FeGe$_2$, enabling us to reproduce the anomalous excitation in the low temperature AFM state. Instead of the one-dimensional (1D) correlation, unexpected two-dimensional (2D) correlations emerge as temperature increase, attributed to the appearance of stripe-type domains and the magnetic frustration. Such 2D correlations are believed to rationalize the anomalous checkerboard-shaped diffusive scattering.

A semi-cylindrical single crystal FeGe$_2$ with an approximate 15 mm radius and 40 mm length with a mass of 110 g was used for the INS measurements. This crystal, used in previous studies of FeGe$_2$ [19,22], was measured using ARCS and HB-3 at Oak Ridge National Laboratory (details in Supplemental Material [23]).

Figure. 1 shows a sample of the INS data acquired as a function of both temperature and energy transfer. Figs. 1(a)-1(b) shows the elastic scattering in the (HK0) planes. At 20 K, one sees the diffraction pattern with several aluminum powder line rings visible (from sample environment background). As the temperature increases, an anomalous feature can be observed clearly in the (HK0) slices of dynamical structure factor S(**Q**, E) (Fig. 1 and Fig. S2). Besides the expected nuclear and magnetic Bragg peaks, there is extra intensity connecting NN magnetic Bragg peaks. This intensity forms a checkerboard arrangement with rods along the X-M directions, as depicted in the 3D rendering at L = 0 (Fig. 1(l)). This checkerboard-shaped anomaly is found in both elastic and inelastic scattering slices and does not strongly depend on energy transfer. At 20 K, such intensity can only be observed at finite energy transfers and appears to be detached from M points, forming a dot-dash-dot pattern (Fig. 1(c)). TAX data at 8 meV shows that each dash consists of two sections, which merge with the magnetic peaks at M points as the temperature increases (Figs. 1(e)-(i)).

The extent of the phonon and magnetic excitation spectra can be assessed by examining the scattering intensity as a function of energy and momentum transfer along the X-M direction for 20 and 300 K as shown in Figs. 2 (a), 2(d) respectively. Magnetic excitations emerge from the M points and disperse up to approximately 30 meV. As momentum transfer increases, the magnetic form factor causes the spin wave scattering intensity to decrease. At larger momentum transfers, one can see the optical phonon excitations between approximately 15 and 35 meV.

Phonon simulations were performed using finite temperature effective force constants (details in Supplemental Material [23]). The simulated phonons match well with experimental data, as shown in Figs. 2(b), 2(e) for X-M direction. The magnetic excitations dispersing out of M points were reproduced as described in Supplemental Material [23] using the NN model, with SJ values from literature [19]. The NN model reproduce some other features of the magnetic spectrum. Along Γ-M direction, the simulated spin wave agrees perfectly with experimental data. Along X-M direction, the agreement is still good in the vicinity of the magnetic zone center, although the measured spin wave appears steeper (Fig. 2(c)).

There are important differences between the measurement and simulation. The main difference is that the continua-like intensity between neighboring M points is not reproduced in either phonon or spin wave simulation. This intensity is associated with the anomalous checkerboard intensity observed in the (HK0) planes. Although over-damped, this intensity can be resolved at 20 K as collective dispersive excitations, which reach the minimums at X points. At 300 K, this dispersive excitation is further damped and merges to a broad response across a wide energy range along rod directions in the Brillouin zone. Spectral weight perpendicular to this direction (Γ-X) is weak and the dispersion is very steep around X points (Fig. S3).

To reveal the origin of the checkerboard anomaly, a **Q**-dependence analysis was performed. It is found this anomalous intensity only appear where L is even (Fig. 1(k)), following the same behavior of magnetic Bragg peaks. This observation indicates that the order behind the anomaly has the same periodicity along the *c* axis as the magnetic structure of $FeGe_2$. To quantify the structural factor of the anomalous INS intensity, equivalent points of [1.3, 0.3, 0] in momentum space were selected on each rod at 0 meV and 8 meV (HK0) slices and their intensity was compared to $AF^2(\mathbf{Q})+B$, where $F(\mathbf{Q})$ is the magnetic form factor of charge-neutral Fe atom, A and B are constants to be fit. Fe atom was used here instead of ions since no charge transfer was found in our charge distribution calculation [24]. Fig. 3 shows the excellent agreement between the fitted curves and the experimental data, further confirming that the anomalous intensity, in both the elastic and inelastic scattering, is indeed from the magnetic origin. At 20 K and 0 meV, as shown in Fig. 3(a), the intensity is dominated by the background, consistent with the fact that no rod intensity was observed in the low temperature elastic scattering. Contributions to the checkerboard anomaly from other sources including electronic scattering or leftover incommensurate order were considered but excluded, as described in Supplemental Material [23].

To describe our observation, a new model is needed. A Checkerboard-shaped anomalous excitation without periodicity along [0, 0, L] has previously been observed in quasi-2D square lattice systems with magnetic frustration, and is described by the $J_1$-$J_2$ Heisenberg model [25,26]. Different from $FeGe_2$, the interlayer interaction $J_c$ is ignored in these systems since it is generally much weaker. Besides the NN interaction $J_1$, the intraplane NNN interaction $J_2$ is also important in describing the excitations in these systems. The ground state of these systems could be determined by the frustration parameter $\eta = J_1/2J_2$, where $J_1$ could be either FM or AFM but $J_2$ is always AFM. Néel-type, stripe-type and FM ordering occurs for $\eta > 1$, $|\eta| < 1$ and $\eta < -1$, respectively. Perfect frustration happens when $|\eta| = 1$. Extreme spatial anisotropy due to the perfect frustration leads to effectively 1D behavior and corresponding plane-like features in INS [27].

Identical to the magnetic atom in these systems, each Fe in $FeGe_2$ has 4 intraplane NN atoms and 4 intraplane NNN atoms. However, $FeGe_2$ has an additional strong interplane interaction $J_c$. To determine the extent of the role of $J_2$ interaction in $FeGe_2$, total energy calculations were made on five collinear magnetic configurations and the exchange parameters were fit to these energies under the assumption that the change of energy is only dependent on the selected exchange interactions (Fig. S5 and Table SI). Two models were used in the least-square fitting: the NN model containing only $J_1$ and $J_c$, and the NNN model that also includes $J_2$. The results are shown in **Error! Reference source not found.**.

For the NN model, the calculations yield $|J_c/J_1| \sim 11.6$, this large ratio represents the anisotropic nature of in-plane and Γ-Z spin waves in $FeGe_2$, consistent with previous report [18]. For the NNN model, $J_1S^2$ and $J_cS^2$ remain almost unchanged from those values determined for the NN model and $J_2S^2$ has a value of -2.6 meV. $J_1$ and $J_2$ are of the same order of magnitude, indicating that both interactions play important roles in describing the spin dynamics and should not be neglected.

Exchange parameters calculated for the NNN model were then used to perform spin wave simulations. Assuming no charge transfer between Fe and Ge atoms and a low-spin configuration in a tetrahedral environment with spin S = 1 and g-factor of 2, the magnetic moment is expected to be $\mu = g\sqrt{S(S+1)}\mu_B = 2.8\ \mu_B$ per Fe atom. However, this is about twice the experimental value, explained as arising from hybridization of the 3d and 4s orbitals of the Fe atoms [14]. As a result, a smaller effective value of S is needed to account for the itinerant nature of $FeGe_2$. A value of 0.5 was chose here so that spin wave along Γ-M matches with result from literature [18]. The simulation in Fig. 2(f) shows the importance of $J_2$ in reducing the spin wave energy near the X points in the Brillouin zone, and significantly improving agreement with the experimental results. A large energy broadening of the spin wave is needed in reproducing the measured spectra, and an additional 10 meV energy broadening is included in the simulation. Such damping of the spin wave may come from the frustration in FeGe$_2$. The simulated (HK0) slice at 8 meV (Fig. 1(j)) shows strong "rod" intensity along X-M and only weak intensity along Γ-X near the X points for a large range of energy transfer, consistent with the experiments.

As temperature increases to 300 K, the dispersive magnetic excitations become more and more diffuse and soften to lower energy transfer. Constant Q cuts were obtained from the TAX data (Figs. S6(a)-S6(e)), revealing a dramatic softening of the spin wave from finite energy at low temperatures to near 0 meV at 300 K as seen in Figs. 4 (a), 4(b). This effect is also visible in Fig. 4(c) showing the constant energy scans at 8 meV with denser temperature points. Between 5 and 275 K, two peaks between neighboring M points can be easily resolved. Above 300 K, the peaks vanish, and the intensity becomes flat between M points. Combining these findings, we believe that the well-defined spin wave collapses near 300 K as one enters the paramagnetic phase of FeGe$_2$.

At the same time as the spin wave collapses, rods of scattering between neighboring M points connect with each other and form the checkerboard-shaped diffusive scattering pattern in (HK0) elastic slices of S(**Q**, E). To gain a quantitative understanding of the spatial coherence behind these rods, correlation lengths were extracted by fitting elastic cuts with a Voigt function across the rod of scattering in the basal plane (Γ-X) and across-plane (Γ-Z), as described in Supplemental Material [23]. Along the rod direction, the almost flat intensity suggests no correlation. At 300 K, the correlation length is about 12 Å in the across-rod direction and 23 Å in the across-plane direction, comparable with each other. In this case, 3D order breaks down with 2D correlations of plates along directions bisecting the *a* and *b* axes remaining in the paramagnetic phase. The 2D correlations unveiled in real space are surprising in FeGe$_2$. In such a quasi-1D system, spins are weakly coupled in the *a-b* plane with fourfold symmetry, and it is expected that correlations in this plane disappear simultaneously with only 1D correlation along *c* axis left. At 500 K, the correlation lengths become 4 Å and 11 Å, respectively, indicating a simultaneous decrease of short-range order upon further warming into the paramagnetic phase.

This striking dimension change can also be explained by $J_2$ and magnetic frustration. Our calculations show that a stripe-type configuration has the next-lowest ground state energy, about 3.7 meV per atom above the Néel-type AFM structure. When the temperature increases to 300 K, thermal fluctuations become comparable to the energy difference between these two configurations and exchange interactions can no longer stabilize the Néel-type structure. It is likely that stripe-type domains start to appear and occupy nearly half of the system. In the Néel-type structure, the effective coupling along both [1, 1, 0] and [1, -1, 0] are of ~ $J_1-2J_2$, which is close to zero when η ~ 1. Once the [1, 1, 0] stripe-type domains are formed, the magnitude of effective coupling along [1, -1, 0] remains nearly zero but that along [1, 1, 0] increases to ~ $J_1+2J_2$. In this case the magnetic order could be viewed as plates perpendicular to the rods along [1, -1, 0], where there are strong in-plate correlations, but the neighboring plates are nearly decoupled because of the small effective correlations between plates. Since [1, 1, 0] and [1, -1, 0] are equivalent directions in the system, stripe-type domains along the other direction are equally likely to be formed, together they can account for the rods in the checkerboard arrangement.

In conclusion, we observe anomalous excitations at low temperature, as well as a checkerboard-shaped diffusive scattering pattern developing at high temperature. We showed that these two phenomena, though different in underling mechanism, both have a magnetic origin and are related to the intraplane NNN interaction $J_2$. This previously ignored interaction generates the extra spin wave feature for a large range of the reciprocal space. Our ab initio calculations show $J_2$ leads directly to the near-perfect in-plane magnetic frustration, which facilitates the emergence of unexpected 2D short-range magnetic order at high temperature. Low dimensional $FeGe_2$ has been synthesized for potential spintronic applications [28,29]. Our revelation of the magnetic frustration and its roles may provide some insights on these studies. Our work also sheds light on the potential of controlling the magnetic dimensionality and corresponding properties of materials by frustration.

**Acknowledgements**

**Funding:** This material is based upon work supported by the National Science Foundation under Grant No. 1750786. The neutron scattering portion of this research used resources at the Spallation Neutron Source and the High Flux Isotope Reactor, both of which are DOE Office of Science User Facilities operated by the Oak Ridge National Laboratory.

**Author Contributions:** D.A. and C.L. performed the INS measurements on ARCS. M.S., M.L. and C.A. performed the INS measurements on HB-3. Y.S. performed the theoretical calculation. Y.S. and C.L. analyzed the data and wrote the manuscript. All authors contributed to discussing and editing the manuscript.

**Competing Interests statement:** The authors declare no competing interests.

**References**

[1]   L. Balents, *Spin Liquids in Frustrated Magnets*, Nature **464**, 199 (2010).

[2]   P. Chandra, P. Coleman, and A. I. Larkin, *Ising Transition in Frustrated Heisenberg Models*, Phys. Rev. Lett. **64**, 88 (1990).

[3]   C. Fang, H. Yao, W. F. Tsai, J. Hu, and S. A. Kivelson, *Theory of Electron Nematic Order in LaFeAsO*, Phys. Rev. B - Condens. Matter Mater. Phys. **77**, 224509 (2008).

[4]   N. Shannon, T. Momoi, and P. Sindzingre, *Nematic Order in Square Lattice Frustrated Ferromagnets*, Phys. Rev. Lett. **96**, 027213 (2006).

[5]   C. Xu, M. Müller, and S. Sachdev, *Ising and Spin Orders in the Iron-Based Superconductors*, Phys. Rev. B - Condens. Matter Mater. Phys. **78**, 020501 (2008).

[6]   B. G. Park, J. Wunderlich, X. Martí, V. Holý, Y. Kurosaki, M. Yamada, H. Yamamoto, A. Nishide, J. Hayakawa, H. Takahashi, A. B. Shick, and T. Jungwirth, *A Spin-Valve-like Magnetoresistance of an Antiferromagnet-Based Tunnel Junction*, Nat. Mater. **10**, 347 (2011).

[7]   P. Wadley, B. Howells, J. Železný, C. Andrews, V. Hills, R. P. Campion, V. Novák, K. Olejník, F. Maccherozzi, S. S. Dhesi, S. Y. Martin, T. Wagner, J. Wunderlich, F. Freimuth, Y. Mokrousov, J. Kuneš, J. S. Chauhan, M. J. Grzybowski, A. W. Rushforth, K. W. Edmonds, B. L. Gallagher, and T. Jungwirth, *Electrical Switching of an Antiferromagnet*, Science (80-. ). **351**, 587 LP (2016).

[8]   A. T. Wong, C. Beekman, H. Guo, W. Siemons, Z. Gai, E. Arenholz, Y. Takamura, and T. Z. Ward, *Strain Driven Anisotropic Magnetoresistance in Antiferromagnetic La 0.4Sr0.6MnO3*, Appl. Phys. Lett. **105**, 052401 (2014).

[9]   Y. Y. Wang, C. Song, B. Cui, G. Y. Wang, F. Zeng, and F. Pan, *Room-Temperature Perpendicular Exchange Coupling and Tunneling Anisotropic Magnetoresistance in an Antiferromagnet-Based Tunnel Junction*, Phys. Rev. Lett. **109**, 137201 (2012).

[10]  T. Jungwirth, X. Marti, P. Wadley, and J. Wunderlich, *Antiferromagnetic Spintronics*, Nat. Nanotechnol. **11**, 231 (2016).

[11] K. Yasukōchi, K. Kanematsu, and T. Ohoyama, *Magnetic Properties of Intermetallic Compounds in Iron-Germanium System : Fe1.67Ge and FeGe2*, J. Phys. Soc. Japan **16**, 429 (1961).

[12] E. Krén and P. Szabó, *Antiferromagnetic Structure of FeGe2*, Phys. Lett. **11**, 215 (1964).

[13] J. B. Forsyth, C. E. Johnson, and P. J. Brown, *The Magnetic Structure and Hyperfine Field of FeGe2*, Philos. Mag. **10**, 713 (1964).

[14] N. S. Satya Murthy, R. J. Begum, C. S. Somanathan, and M. R. L. N. Murthy, *Magnetic Structures in the Iron-Germanium System*, Solid State Commun. **3**, 113 (1965).

[15] E. F. Bertaut and J. Chenavas, *Comments on the Magnetic Structure of ∂-FeGe2*, Solid State Commun. **3**, 117 (1965).

[16] J. Sólyom and E. Krén, *On the Magnetic Structure of FeGe2*, Solid State Commun. **4**, 255 (1966).

[17] V. V Tarasenko, V. Pluzhnikov, and E. Fawcett, *Theory of the Magnetic Phase Diagram and Magnetostriction of FeGe2*, Phys. Rev. B **40**, 471 (1989).

[18] T. M. Holden, A. Z. Menshikov, and E. Fawcett, *Anisotropic Spin-Wave Dispersion in FeGe2*, J. Phys. Condens. Matter **8**, L291 (1996).

[19] C. P. Adams, T. E. Mason, E. Fawcett, A. Z. Menshikov, C. D. Frost, J. B. Forsyth, T. G. Perring, and T. M. Holden, *High-Energy Magnetic Excitations and Anomalous Spin-Wave Damping in FeGe2*, J. Phys. Condens. Matter **12**, 8487 (2000).

[20] H. Zhou, *Room Temperature Section of the Phase Diagram of the Cu-Fe-Ge Ternary System*, J. Less-Common Met. **171**, 113 (1991).

[21] L. M. Corliss, J. M. Hastings, W. Kunnmann, R. Thomas, J. Zhuang, R. Butera, and D. Mukamel, *Magnetic Structure and Critical Properties of FeGe2*, Phys. Rev. B **31**, 4337 (1985).

[22] H. L. Smith, Y. Shen, D. S. Kim, F. C. Yang, C. P. Adams, C. W. Li, D. L. Abernathy, M. B. Stone, and B. Fultz, *Temperature Dependence of Phonons in FeGe2*, Phys. Rev. Mater. **2**, 103602 (2018).

[23] See Supplemental Material at [URL Will Be Inserted by Publisher] for Details on the Material Structure, Experimental Details, More INS Data, Phonon and Spin Wave Simulation, Fermi Surface, Exchange Parameters Calculation and Correlation Length Fittings.

[24] G. Henkelman, A. Arnaldsson, and H. Jónsson, *A Fast and Robust Algorithm for Bader Decomposition of Charge Density*, Comput. Mater. Sci. **36**, 354 (2006).

[25] N. Shannon, B. Schmidt, K. Penc, and P. Thalmeier, *Finite Temperature Properties and Frustrated Ferromagnetism in a Square Lattice Heisenberg Model*, Eur. Phys. J. B **38**, 599 (2004).

[26] D. C. Johnston, *The Puzzle of High Temperature Superconductivity in Layered Iron Pnictides and Chalcogenides*, Advances in Physics.

[27] A. Sapkota, B. G. Ueland, V. K. Anand, N. S. Sangeetha, D. L. Abernathy, M. B. Stone, J. L. Niedziela, D. C. Johnston, A. Kreyssig, A. I. Goldman, and R. J. McQueeney, *Effective One-Dimensional Coupling in the Highly Frustrated Square-Lattice Itinerant Magnet CaCo2-YAs2*, Phys. Rev. Lett. **119**, (2017).

[28] D. Czubak, S. Gaucher, L. Oppermann, J. Herfort, K. Zollner, J. Fabian, H. T. Grahn, and M. Ramsteiner, *Electronic and Magnetic Properties of α-FeGe2 Films Embedded in Vertical Spin Valve Devices*, Phys. Rev. Mater. **4**, 104415 (2020).

[29] S. Tang, I. Kravchenko, T. Z. Ward, Q. Zou, J. Yi, C. Ma, M. Chi, G. Cao, A.-P. Li, D. Mandrus, and Z. Gai, *Dimensionality Effects in FeGe2 Nanowires: Enhanced Anisotropic Magnetization and Anomalous Electrical Transport*, Sci. Rep. **7**, 7126 (2017).

**Figures and Tables**

**FIG. 1. Checkerboard-shaped anomaly is observed in the (HK0) planes.**

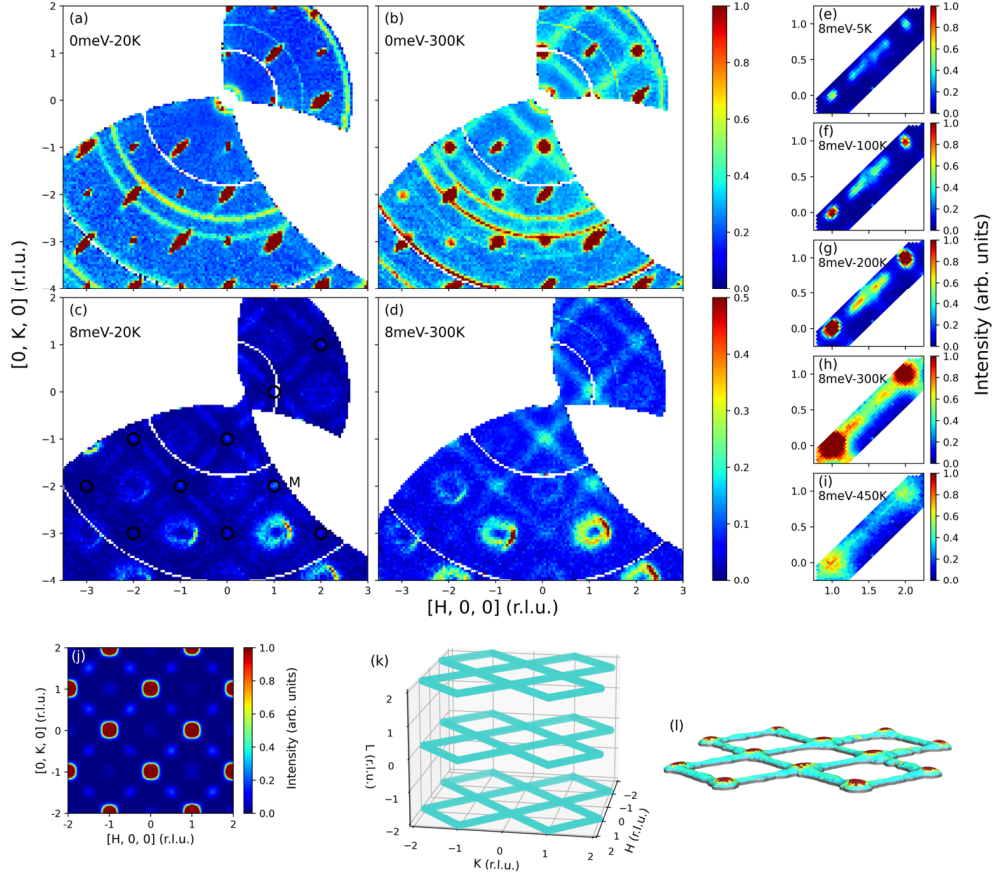

**FIG. 1. Checkerboard-shaped anomaly is observed in the (HK0) planes.** (a)-(d): S(**Q**, E) slices from ARCS measurements, obtained by integrating [-0.5, 0.5] meV in energy and [-0.1, 0.1] reciprocal lattice units (r.l.u.) along [0, 0, L]. (a)(b): Elastic scattering results, (c)(d): inelastic scattering results with a neutron energy loss of 8 meV. In (a)(b), those peaks with H+K even are nuclear Bragg peaks (Γ), and those with H+K odd are magnetic Bragg peaks (M). In general, the intensities of nuclear Bragg peaks are stronger than that of magnetic ones and will increase with respect to the absolute value of momentum transfer |**Q**|. The intensities of magnetic Bragg peaks are, on the other hand, weaker at larger |**Q**|. (e)-(i): S(**Q**, E) slices from TAX measurements, which is more limited but has better resolution. (j): Spin wave simulation of (HK0) plane at 8 meV from NNN model. (k): Three-dimensional (3D) schematic of the checkerboard-shaped anomaly. (l): 3D rendering of L=0 schematic.

**FIG. 2. Dispersive scattering intensity in several Brillouin zones in slices along [H, H-1, 0].**

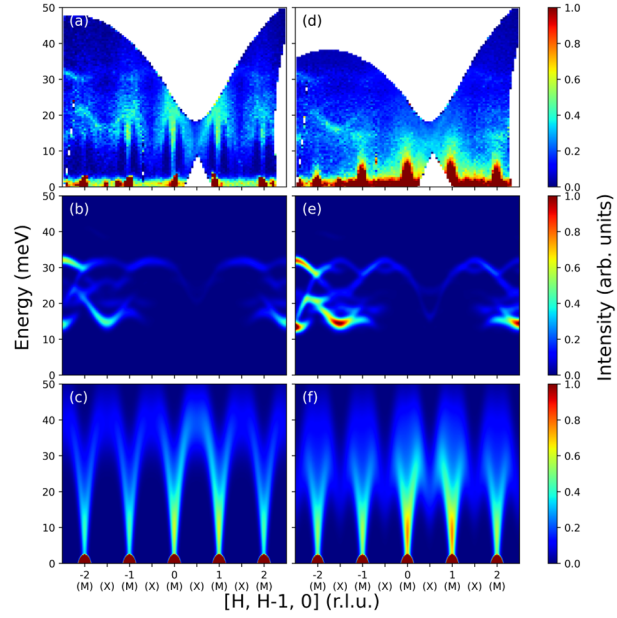

**FIG. 2. Dispersive scattering intensity in several Brillouin zones in slices along [H, H-1, 0].** (a)(b) 20K. (d)(e) 300K. (a) and (d) are S(**Q**, E) slices from ARCS measurements, obtained by integrating [-0.1, 0.1] r.l.u. along [0, 0, L] and [-0.05, 0.05] r.l.u. along [H, -H, 0]. (b)(e) are simulated phonons. (c)(f) are simulated spin waves from the NN model and NNN model.

**FIG. 3. The |Q|-dependence of the checkerboard-shaped anomaly matches well with magnetic form factor of Fe.**

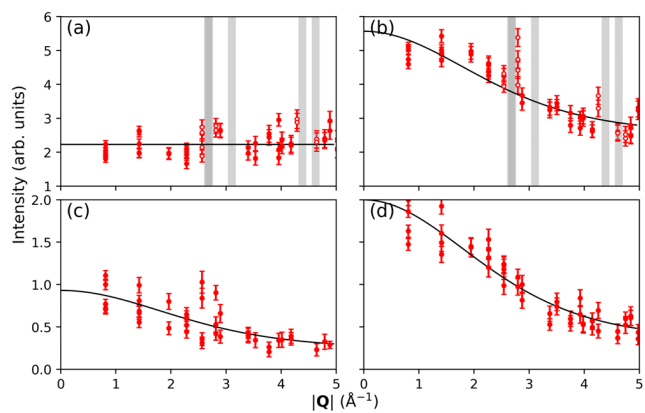

**FIG. 3. The |Q|-dependence of the checkerboard-shaped anomaly matches well with magnetic form factor of Fe.** The fittings were done on 20 K (a)(c) and 300 K (b)(d) ARCS data for energy transfer of 0 meV (a)(b) and 8 meV (c)(d). The red solid circles are experimental data used for fitting, and the black solid lines are the fitted results. The gray vertical stripes in (a)(b) are places where data points were excluded (red open circles) due to the sample environment background.

**FIG. 4. Temperature dependence of the anomalous intensity along [H, H-1, 0].**

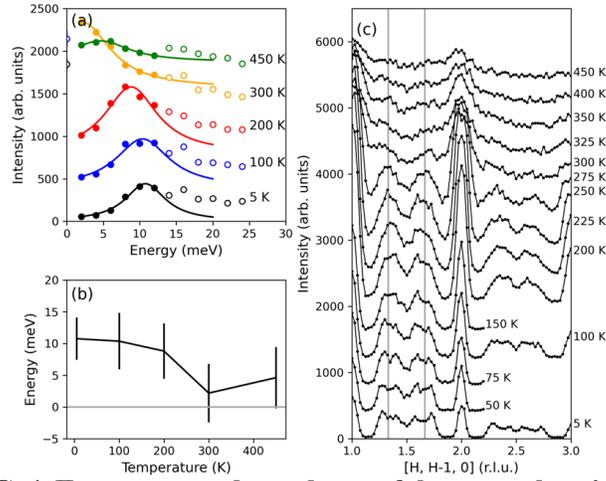

**FIG. 4. Temperature dependence of the anomalous intensity along [H, H-1, 0].** (a) Constant Q cuts at [1.6, 0.6, 0] from TAX data. (b) Lorentzian fit to the first peak (solid circles) in (a), error bars represent the full width at half maximum. (c) Constant E scans along [H, H-1, 0] at 8 meV with the black solid lines as guides to the eye. The vertical gray lines indicate peaks at about 1 and 2 r.l.u.. Error bars are smaller than the symbols.

**Table I. Parameters obtained from total energy calculations.**

| Model | $J_1S^2$ (meV) | $J_2S^2$ (meV) | $J_cS^2$ (meV) | $\eta$ |
|---|---|---|---|---|
| NN | -7.2 ± 2.2 | - | 83 ± 4 | - |
| NNN | -7.2 ± 0.5 | -2.6 ± 0.4 | 81.1 ± 0.9 | 1.4 |

# Supplemental Material

# Frustration-induced diffusive scattering anomaly and dimension change in FeGe$_2$

Yaokun Su[1], Hillary L. Smith[2], Matthew B. Stone[3], Douglas L. Abernathy[3], Mark D. Lumsden[3], Carl P. Adams[4], and Chen Li[1,5*]

1. Materials Science and Engineering, University of California, Riverside, Riverside, CA 92521, USA
2. Physics and Astronomy, Swarthmore College, Swarthmore, PA 19081, USA
3. Neutron Scattering Division, Oak Ridge National Laboratory, Oak Ridge, TN 37831, USA
4. Department of Physics, St. Francis Xavier University, Antigonish, Nova Scotia B2G 2W5, Canada
5. Mechanical Engineering, University of California, Riverside, Riverside, CA 92521, USA

*chenli@ucr.edu

**This file includes:**

Supplemental Text

Figs. S1 to S6

Tables SI to SII

**Supplemental Text**

Crystal structure and Brillouin zone map of FeGe$_2$

FeGe$_2$ has the same body-centered tetragonal crystal structure (Fig. S1) as $\theta$-phase Al$_2$Cu with the space group *I4/mcm*. It exhibits two zero-field magnetic phase transitions on heating: one first-order transition from a commensurate AFM state to an incommensurate spin-density-wave state at 263 K, and another second-order Néel transition from the incommensurate state to paramagnetic phase at 289 K. (HK0) slices were defined such that at the M points **Q** = [100] and at the X points **Q** = $[\frac{1}{2}\frac{1}{2}0]$.

Experimental details

The majority of the data presented were obtained using the time-of-flight Wide Angular Range Chopper Spectrometer (ARCS) [1] at the Spallation Neutron Source (SNS). The sample was mounted inside a low-background electrical resistance vacuum furnace with the (HK0) crystallographic plane horizontal. The measurements were performed with an incident neutron energy of 70 meV, at 20, 300, 500 and 635 K. The crystal was rotated about the vertical axis through 120° in increments of 0.5°. Temperature dependence of phonons in FeGe$_2$ were previously reported [2].

Data reduction and analysis of the ARCS data were performed with MANTID [3]. The data were normalized by the proton current on the spallation target. Bad detector pixels were identified and masked, and the data were corrected for detector efficiency using a measurement of a vanadium standard. After data reduction, neutron events at different detectors were combined to generate the four-dimensional dynamical structure factor S(**Q**, E), where **Q** is the neutron momentum transfer and E is the energy transfer. Two-dimensional slices and one-dimensional cuts were then be obtained by integrating portions of S(**Q**, E) over small ranges.

Additional data were collected with the Triple-Axis Spectrometer (TAX) HB-3 at the High Flux Isotope Reactor (HFIR). The measurements were made between 5 and 450 K with a fixed final energy of 14.7 meV. A pyrolytic graphite (002) monochromator and an analyzer were used, and the horizontal spectrometer collimation was set to 48'-40'-40'-120'. The sample was aligned roughly in the (HHL) scattering plane.

Anomalous magnetic scattering in the (HK0) planes

An anomalous feature can be observed in the (HK0) slices of dynamical structure factor S(**Q**, E), as shown in Fig. S2. The anomalous intensity is observed at 20 K in finite energy transfer up to 25 meV. At 300 K, a checkerboard-shaped pattern was found from 0 to 20 meV. The anomaly weakens but still persist at 500 K and finally vanishes at 635 K as magnetic order collapses.

Finite temperature phonon simulation

Phonopy [4] was used to simulate phonon S(**Q**, E) from the finite temperature effective force constants. The energy resolution of ARCS and a momentum resolution of 0.1 r.l.u. was convoluted in the phonon simulation.

The finite-temperature force constants were calculated with the TDEP method [5–7]. In this procedure, the Born-Oppenheimer surface of FeGe$_2$ at 20 and 300 K was sampled using molecular dynamics (MD) implemented in the VASP package [8–10]. Then, the model Hamiltonian

$$\widehat{H} = U_0 + \sum_i \frac{P_i^2}{2m_i} + \frac{1}{2!}\sum_{ij}\sum_{\alpha\beta}\Phi_{ij}^{\alpha\beta}u_i^\alpha u_j^\beta$$

was fit with this surface, which is done by comparing the forces of the model and the MD at each time step and minimizing the difference. Here, $U_0$ is the potential energy and $\Phi_{ij}$ and $\Phi_{ijk}$ are the second- and third-order force constants. The displacement of atom $i$ from ideal positions is denoted as $\boldsymbol{u}_i$, its momentum $\boldsymbol{p}_i$, and $\alpha, \beta, \gamma$ are Cartesian indices.

For MD calculations, the projector augmented wave (PAW) pseudopotentials [11] were used with the Perdew-Burke-Ernzerhof (PBE) exchange-correlation functional [12]. The plane-wave cutoff was 630 eV. A supercell of a 2 × 2 × 2 conventional unit cell with 96 atoms was used with a 4 × 4 × 4 k-point grid. Lattice constants were relaxed before MD to $a$ = 5.914 Å and $c$ = 4.948 Å, which are slightly larger than the results from neutron powder diffraction at both temperatures [2]. The overestimation of lattice constants in calculation was believed to account for the overall softer phonons in simulation and was corrected by applying a scaling factor of 1.07 for all the modes.

Spin wave simulation

The SpinW program [13] was used to simulate the spin wave and corresponding S(**Q**, E) using classical linear spin wave theory for both the NN model and NNN model. The energy resolution (ARCS resolution plus an additional 10 meV broadening for reproducing damping) and a momentum resolution of 0.1 r.l.u. was convoluted in the S(**Q**, E) simulation. It should be noted that along Γ-Z the spin wave is very steep, so even small integration range in making slices from ARCS S(**Q**, E) results in the wide linewidth of excitations. And this has been reproduced by using the same integration range in the spin wave simulations as the experimental data.

Across-rod spectra

Spectral weight perpendicular to the rods (Γ-X) is weak and the dispersion is very steep around X points (Fig. S3), revealing great anisotropy of this excitation.

Contributions to the checkerboard anomaly from other sources

Contributions to the checkerboard anomaly from other sources including electronic scattering or leftover incommensurate order were considered but excluded.

An electronic scattering process, such as observed in CePd$_3$ [14], was considered. The Fermi surface was calculated with a non-magnetic primitive cell. A 10 × 10 × 10 k-point grid was used in the calculation. There are three bands across Fermi surface, shown in different colors (Fig. S4). To produce the checkerboard-shaped pattern in elastic scattering, certain scattering channels for electrons should exist, which require specific shape of the Fermi surface. However, the condition is not met for all three bands after exploring the data.

The intensity anomaly cannot be explained by leftover incommensurate order either. The magnetic satellites intensity from the incommensurate state, if exists, will be along different directions in reciprocal space.

Besides, the maximum value of δ is 0.05, which is too small to account for the anomaly far away from M points.

Exchange parameters calculation

Exchange constants were calculated using atomistic simulation based on density functional theory. Total energy calculations were made on five collinear magnetic configurations and the exchange parameters were fit to these energies under the assumption that the change of energy is only dependent on the exchange interactions (Fig. S5 and Table SI). $E_0$ represents the part of total energy that are independent of exchange interactions. The plane-wave cutoff was 850 eV. A supercell of a 2 × 2 × 2 conventional unit cell with 96 atoms was used with a 12 × 12 × 12 k-point grid. Results are shown in main text.

$$E = E_0 - N_1 J_1 S^2 - N_c J_c S^2$$

$$E = E_0 - N_1 J_1 S^2 - N_2 J_2 S^2 - N_c J_c S^2$$

It should be noted that a stripe-type configuration has the next-lowest ground state energy. At 300 K, the energy difference between stripe-type and Néel-type configurations is about 3.7 meV per atom and the $k_B T$ becomes 25.8 meV, so the ratio between these two configurations $N_3/N_1 = e^{-\Delta E/k_B T} \approx 0.86$.

Correlation length fittings

To gain a quantitative understanding of the spatial coherence behind these rods, correlation lengths were extracted by fittings elastic cuts with a Voigt function across the rod of scatting in the basal plane (Γ-X) and across-plane (Γ-Z) (Table SII). Along the rod direction, the almost flat intensity suggests no correlation.

The Voigt function is of the form:

$$I(q) \propto A \int_{-\infty}^{\infty} e^{-\frac{\tau^2}{2\sigma^2}} \frac{\gamma}{(q-\tau)^2 + \gamma^2} d\tau$$

Where A is amplitude, σ is the standard deviation determined by fitting a nearby nuclear Bragg peak to a Gaussian function, γ = 1/ξ is the inverse of the real-space correlation length and q is the wave vector away from the peak center.

**References**

[1]  D. L. Abernathy, M. B. Stone, M. J. Loguillo, M. S. Lucas, O. Delaire, X. Tang, J. Y. Y. Lin, and B. Fultz, *Design and Operation of the Wide Angular-Range Chopper Spectrometer ARCS at the Spallation Neutron Source*, Review of Scientific Instruments.

[2]  H. L. Smith, Y. Shen, D. S. Kim, F. C. Yang, C. P. Adams, C. W. Li, D. L. Abernathy, M. B.

Stone, and B. Fultz, *Temperature Dependence of Phonons in FeGe2*, Phys. Rev. Mater. **2**, 103602 (2018).

[3] O. Arnold, J. C. Bilheux, J. M. Borreguero, A. Buts, S. I. Campbell, L. Chapon, M. Doucet, N. Draper, R. Ferraz Leal, M. A. Gigg, V. E. Lynch, A. Markvardsen, D. J. Mikkelson, R. L. Mikkelson, R. Miller, K. Palmen, P. Parker, G. Passos, T. G. Perring, P. F. Peterson, S. Ren, M. A. Reuter, A. T. Savici, J. W. Taylor, R. J. Taylor, R. Tolchenov, W. Zhou, and J. Zikovsky, *Mantid - Data Analysis and Visualization Package for Neutron Scattering and µ SR Experiments*, Nucl. Instruments Methods Phys. Res. Sect. A Accel. Spectrometers, Detect. Assoc. Equip. **764**, 156 (2014).

[4] A. Togo and I. Tanaka, *First Principles Phonon Calculations in Materials Science*, Scr. Mater. **108**, 1 (2015).

[5] O. Hellman, I. A. Abrikosov, and S. I. Simak, *Lattice Dynamics of Anharmonic Solids from First Principles*, Phys. Rev. B - Condens. Matter Mater. Phys. **84**, 180301 (2011).

[6] O. Hellman and I. A. Abrikosov, *Temperature-Dependent Effective Third-Order Interatomic Force Constants from First Principles*, Phys. Rev. B - Condens. Matter Mater. Phys. **88**, 144301 (2013).

[7] O. Hellman, P. Steneteg, I. A. Abrikosov, and S. I. Simak, *Temperature Dependent Effective Potential Method for Accurate Free Energy Calculations of Solids*, Phys. Rev. B - Condens. Matter Mater. Phys. **87**, 104111 (2013).

[8] G. Kresse and J. Hafner, *Ab Initio Molecular Dynamics for Liquid Metals*, Phys. Rev. B **47**, 558 (1993).

[9] G. Kresse and J. Furthmüller, *Efficiency of Ab-Initio Total Energy Calculations for Metals and Semiconductors Using a Plane-Wave Basis Set*, Comput. Mater. Sci. **6**, 15 (1996).

[10] G. Kresse and J. Furthmüller, *Efficient Iterative Schemes for Ab Initio Total-Energy Calculations Using a Plane-Wave Basis Set*, Phys. Rev. B - Condens. Matter Mater. Phys. **54**, 11169 (1996).

[11] D. Joubert, *From Ultrasoft Pseudopotentials to the Projector Augmented-Wave Method*, Phys. Rev. B - Condens. Matter Mater. Phys. **59**, 1758 (1999).

[12] J. P. Perdew, K. Burke, and M. Ernzerhof, *Generalized Gradient Approximation Made Simple*, Phys. Rev. Lett. **77**, 3865 (1996).

[13] S. Toth and B. Lake, *Linear Spin Wave Theory for Single-Q Incommensurate Magnetic Structures*, J. Phys. Condens. Matter **27**, 10 (2015).

[14] E. A. Goremychkin, H. Park, R. Osborn, S. Rosenkranz, J.-P. Castellan, V. R. Fanelli, A. D. Christianson, M. B. Stone, E. D. Bauer, K. J. McClellan, D. D. Byler, and J. M. Lawrence, *Coherent Band Excitations in CePd3: A Comparison of Neutron Scattering and Ab Initio Theory*, Science (80-. ). **359**, 186 LP (2018).

[15] M. I. Aroyo, D. Orobengoa, G. De La Flor, E. S. Tasci, J. M. Perez-Mato, and H. Wondratschek, *Brillouin-Zone Database on the Bilbao Crystallographic Server*, Acta Crystallogr. Sect. A Found. Adv. **70**, 126 (2014).

[16] A. Kokalj, *XCrySDen-a New Program for Displaying Crystalline Structures and Electron Densities*, J. Mol. Graph. Model. **17**, 176 (1999).

**Fig. S1.**

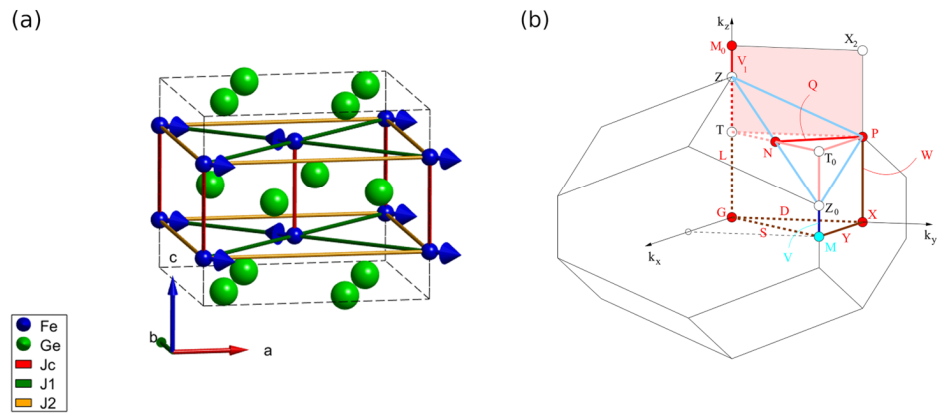

Fig. S1. Crystal structure and Brillouin zone map of FeGe2. (a) Crystal structure of $FeGe_2$ with Neel-type AFM magnetic configuration. Exchange interactions $J_c$, $J_1$ and $J_2$ are labeled in the figure. (b) The Brillouin zone map of $FeGe_2$. [15]

**Fig. S2.**

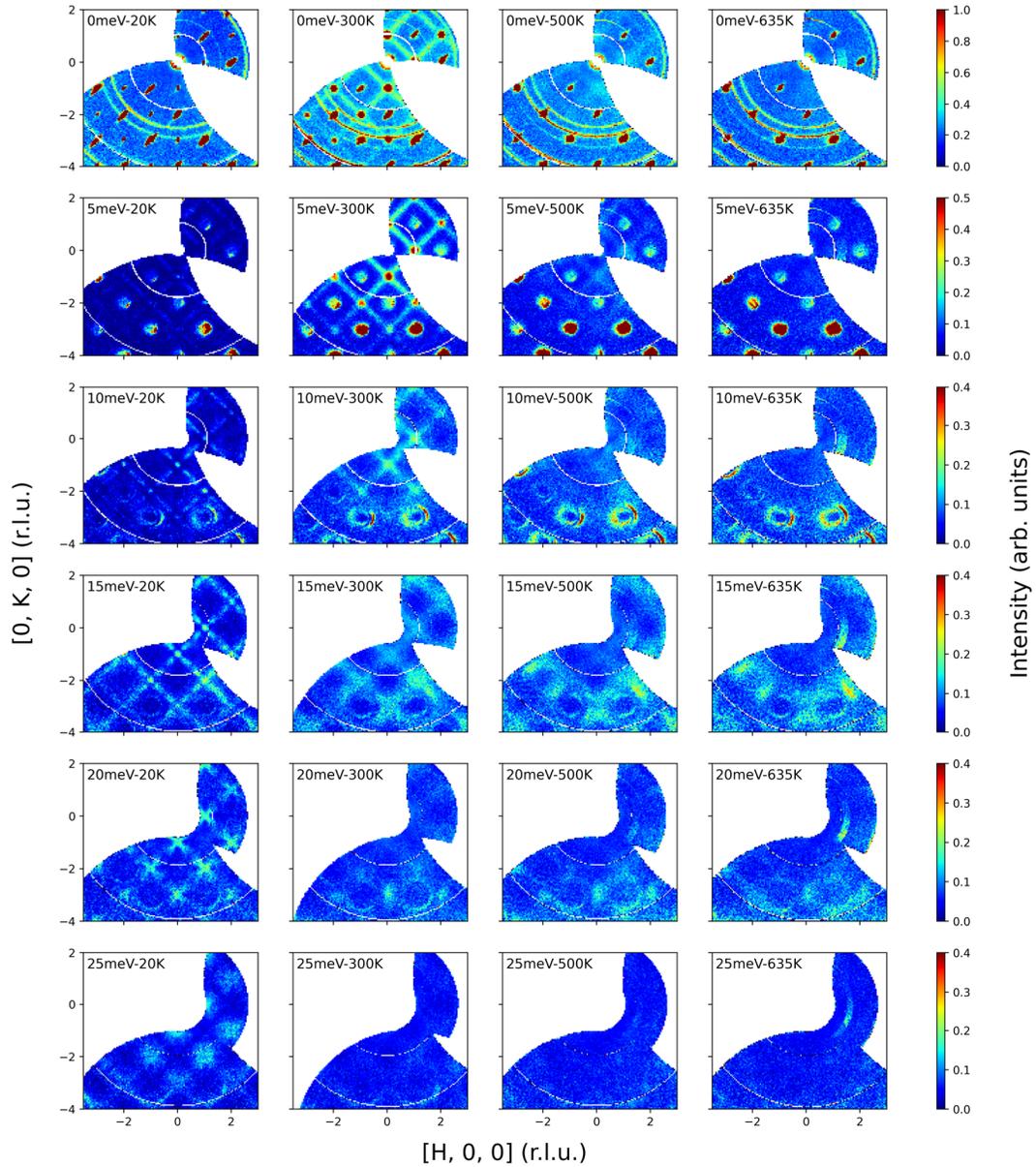

Fig. S2. Checkerboard-shaped anomaly is observed in the (HK0) planes from ARCS measurements. Slices were obtained by integrating from -0.5 to 0.5 meV in energy and from -0.1 to 0.1 reciprocal lattice units (r.l.u.) along [0, 0, L].

**Fig. S3.**

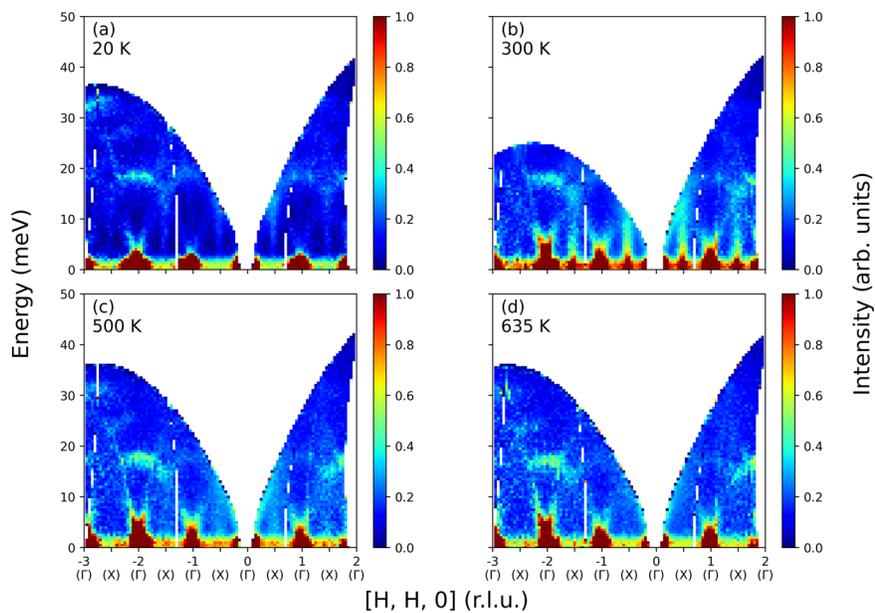

Fig. S3. Across-rod spectra in several Brillouin zones in slices along [H, H, 0]. The anomalous intensity is very steep and only exist near X points (half-integer H values). Slices were obtained by integrating from -0.1 to 0.1 r.l.u. along [0, 0, L] and from -0.05 to 0.05 r.l.u. along [H, -H, 0].

**Fig. S4.**

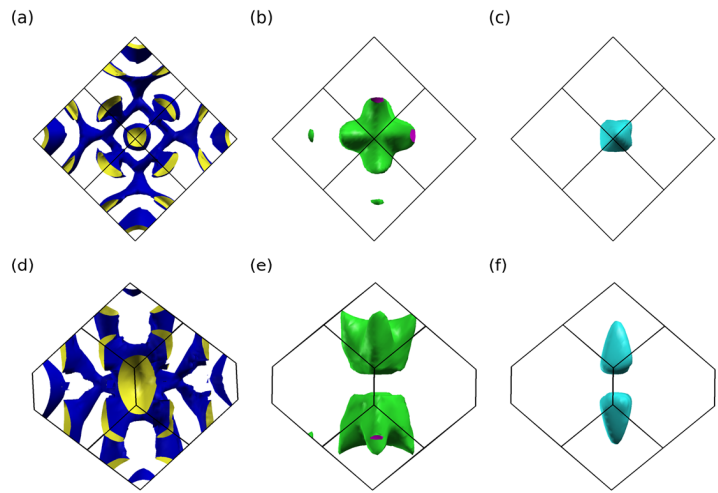

Fig. S4. Fermi surface of non-magnetic FeGe$_2$. (a)(b)(c) Top views. (d)(e)(f) Side views. The results are visualized by XCrySDen [16].

**Fig. S5.**

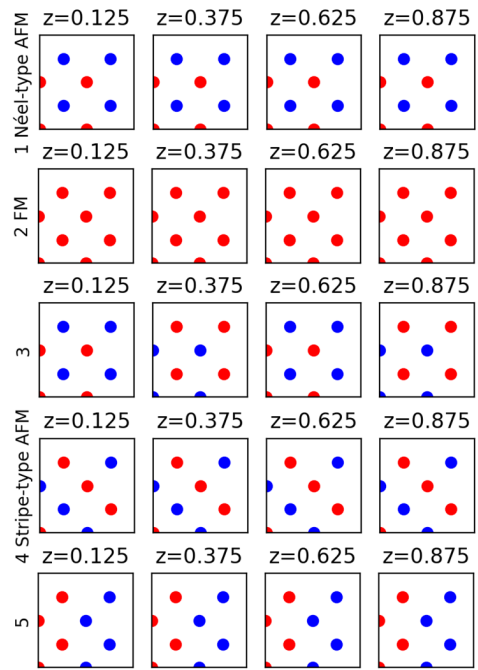

Fig. S5. Configurations used for exchange parameters calculation. Red and blue circles represent Fe atoms with spin up and spin down. Each configuration is identified by the arrangement of Fe atoms in four a-b planes stacked along c axis.

**Fig. S6.**

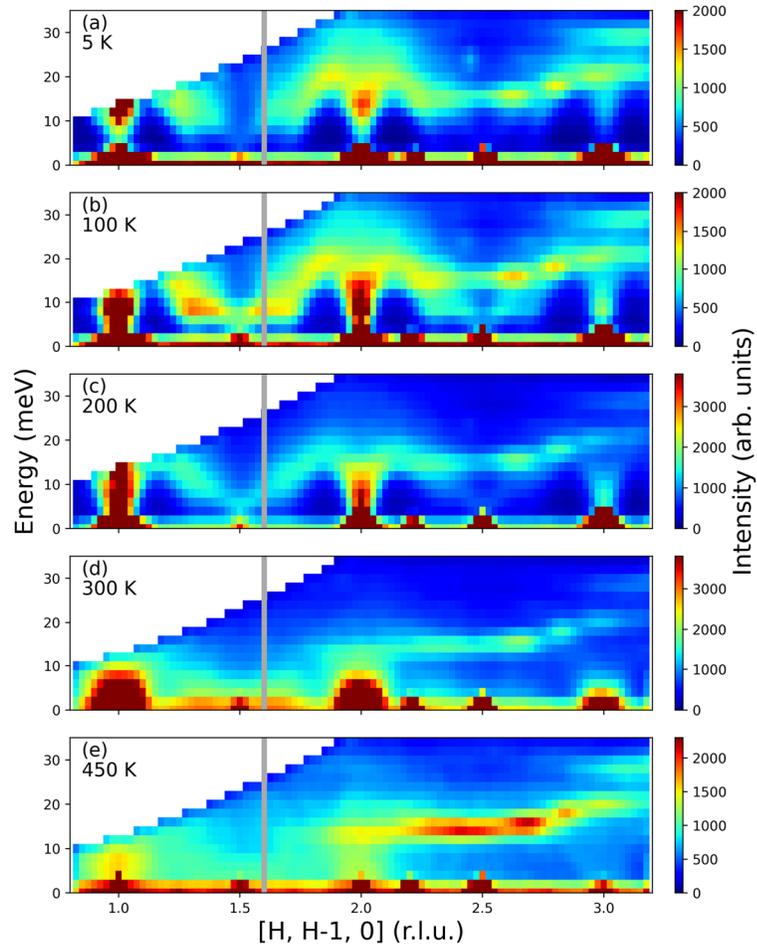

Fig. S6. Temperature dependence of the anomalous intensity along [H, H-1, 0]. (a)-(e): S(**Q**, E) slices from TAX measurements at 5, 100, 200, 300 and 450 K, respectively. The gray vertical stripes indicate the locations of the constant-Q cuts in main text.

**Table SI. Total energy of five magnetic configurations.**

| Configuration | Energy (eV) |
|---|---|
| 1 | -560.1304 |
| 2 | -559.2104 |
| 3 | -554.9302 |
| 4 | -560.0107 |
| 5 | -559.7923 |

**Table SII. Correlation lengths extracted along across-rod and across-plane directions.**

| Temperature (K) | Across-rod $\xi$ (Å) | Across-plane $\xi$ (Å) |
|---|---|---|
| 300 | 11.6 ± 0.9 | 23 ± 2 |
| 500 | 4.2 ± 0.5 | 11 ± 1 |